\documentclass{ws-ijmpc}

\begin{document}

\markboth{H.-X. Yang $et$ $al.$} {Traffic-driven epidemic spreading
on scale-free networks with tunable degree distribution}

\catchline{}{}{}{}{}

\title{Traffic-driven epidemic spreading
on scale-free networks with tunable degree distribution}

\author{Han-Xin Yang\footnote{hxyang01@gmail.com}}

\address{Department of Physics, Fuzhou University \\Fuzhou
350108, China}

\author{Bing-Hong Wang\footnote{bhwang@ustc.edu.cn}}
\address{Department of Modern Physics, University of
Science and Technology of China \\Hefei, 230026, China}

\maketitle

\begin{history}

\end{history}

\begin{abstract}
We study the traffic-driven epidemic spreading on scale-free
networks with tunable degree distribution. The heterogeneity of
networks is controlled by the exponent $\gamma$ of power-law degree
distribution. It is found that the epidemic threshold is minimized
at about $\gamma=2.2$. Moreover, we find that nodes with larger
algorithmic betweenness are more likely to be infected. We expect
our work to provide new insights into the effect of network
structures on traffic-driven epidemic spreading.

\end{abstract}

\ccode{PACS Nos.: 89.75.Hc, 89.75.Fb}

\section{Introduction}

Epidemic spreading~\cite{1,2,3,4,5,6,7,8,9,10,11,12} and traffic
transportation~\cite{12,13,14,15,16,17,18,19,19.1,20,21,22} on
complex networks have attracted much attention in the past decade.
In many cases, epidemic spreading is relied on the process of
transportation. For example, a computer virus can spread over
Internet via data transmission. Another example is that air
transport tremendously accelerates the propagation of infectious
diseases among different countries.

The first attempt to incorporate traffic into epidemic spreading is
based on metapopulation model~\cite{m1,m2,m3,m4,m5,m6}. This
framework describes a set of spatially structured interacting
subpopulations as a network, whose links denote the traveling path
of individuals across different subpopulations. Each subpopulation
consists of a large number of individuals. An infected individual
can infect other individuals in the same subpopulation. In a recent
work, Meloni $et$ $al.$ proposed another traffic-driven epidemic
spreading model~\cite{Meloni}, in which each node of a network
represents a router in the Internet and the epidemic can spread
between nodes by the transmission of packets. A susceptible node
will be infected with some probability every time it receives a
packet from an infected neighboring node.

The routing strategy plays an important role in the traffic-driven
epidemic spreading. Meloni $et$ $al$. observed that when travelers
decide to avoid locations with high levels of prevalence, this
self-initiated behavioral change may enhance disease
spreading~\cite{avoid}. Yang $et$ $al$. found that epidemic
spreading can be effectively controlled by a local routing
strategy~\cite{yang1}, a greedy routing~\cite{yang2} or an efficient
routing protocol~\cite{yang3}. For a given routing strategy, the
traffic-driven epidemic spreading is affected by network structures.
It has been found that the increase of the average network
connectivity can slow down the epidemic outbreak~\cite{yang4}.
Besides, the epidemic threshold can be enhanced by the targeted
cutting of links among large-degree nodes or edges with the largest
algorithmic betweenness~\cite{yang5}.

Many real networks display an power-law degree distribution:
$P(k)\sim k^{-\gamma}$, with the exponent typically satisfying
$2<\gamma \leq3$~\cite{exponent}. It has been found that the
exponent of power-law degree distribution plays an important role in
opinion dynamics~\cite{opinion1,opinion2} and evolutionary
games~\cite{yang6}. In this paper, we study how the exponent of
power-law degree distribution affects the traffic-driven epidemic
spreading. Our preliminary results have shown that there exists an
optimal value of exponent, leading to the minimum epidemic
threshold.

The paper is organized as follows. In Sec.~\ref{sec:network}, we
introduce scale-free networks with tunable degree distribution. In
Sec.~\ref{sec:model}, we describe the traffic-driven epidemic
spreading model. The results and discussions are presented in
Sec.~\ref{sec:results}. Finally, we give a brief conclusion in
Sec.~\ref{sec:conclusion}.

\section{Scale-free networks with tunable degree distribution}\label{sec:network}

We adopt the algorithm proposed by Dorogovtsev $et$
$al$.~\cite{Dorogovtsev} to generate the scale-free networks with
tunable degree distribution.

Initially, there are $m$ fully connected nodes. At each time, a
newly added node makes $m$ links to $m$ different nodes already
present in the network. The probability $\Pi_{i}$ that the new node
will be connected to an old node $i$ is:

\begin{equation}
\Pi_{i}=\frac{k_{i}+Am}{\Sigma_{j}(k_{j}+Am)},\label{1}
\end{equation}
where $k_{i}$ is the degree of node $i$, the sum runs over all old
nodes, and $A$ is a tunable parameter ($A>-1$). After a long
evolution time, this algorithm generates a scale-free network
following the power-law degree distribution $P(k)\sim k^{-\gamma}$
with the degree exponent $\gamma=3+A$. Particularly, this algorithm
produces the Barabasi-Albert network model~\cite{ba} when $A=0$. The
average degree of the network $\langle k \rangle=2m$.

\section{Traffic-driven epidemic spreading model}\label{sec:model}

Following the work of Meloni $et$ $al.$~\cite{Meloni}, we
incorporate the traffic dynamics into the classical
susceptible-infected-susceptible model~\cite{SIS} of epidemic
spreading as follows.

In a network of size $N$, at each time step, $\lambda N$ new packets
are generated with randomly chosen sources and destinations (we call
$\lambda$ as the packet-generation rate), and each node can deliver
at most $C$ packets towards their destinations. Packets are
forwarded according to a given routing algorithm. The queue length
of each agent is assumed to be unlimited. The first-in-first-out
principle applies to the queue. Each newly generated packet is
placed at the end of the queue of its source node. Once a packet
reaches its destination, it is removed from the system. After a
transient time, the total number of delivered packets at each time
will reach a steady value, then an initial fraction of nodes
$\rho_{0}$ is set to be infected (we set $\rho_{0}=0.1$ in numerical
experiments). The infection spreads in the network through packet
exchanges. Each susceptible node has the probability $\beta$ of
being infected every time it receives a packet from an infected
neighbor. The infected nodes recover at rate $\mu$ (we set $\mu=1$
in this paper).

\section{Results and discussions}\label{sec:results}

\begin{figure}
\begin{center}
\scalebox{0.45}[0.45]{\includegraphics{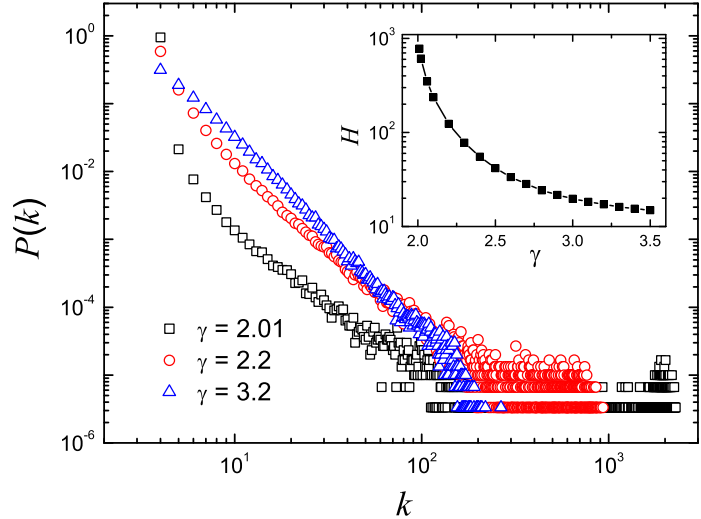}} \caption{(Color
online) The degree distribution $P(k)$ for different values of
$\gamma$. The inset shows the degree heterogeneity $H$ as a function
of $\gamma$. Each data results from an average over 100 different
network realizations.} \label{fig1}
\end{center}
\end{figure}

In the following, we carry out simulations systematically by
employing traffic-driven epidemic spreading on scale-free networks
with tunable degree distribution. We set the size of the network $N
= 3000$ and the packet-generation rate $\lambda=0.5$. Moreover, we
assume that the node-delivering capacity $C$ is infinite, so that
traffic congestion will not occur in the network. Packets are
forwarded according to the shortest-path routing protocol.

Figure~\ref{fig1} shows the degree distribution $P(k)$ for different
values of $\gamma$. One can see that there are more small-degree
nodes and less large-degree nodes in network as $\gamma$ increases.
Following Ref.~\cite{perc}, we quantify the degree heterogeneity of
a network as
\begin{equation}
H= \frac{\langle k ^{2}\rangle-\langle k \rangle}{\langle k
\rangle}.\label{2}
\end{equation}
From the inset of Fig.~\ref{fig1}, one can see that the degree
heterogeneity $H$ decreases as $\gamma$ increases, indicating that
the generated network becomes more homogeneous for larger exponent.

Figure~\ref{2} shows the density of infected nodes $\rho$ as a
function of the spreading rate $\beta$ for different values of the
exponent $\gamma$. One can observe that there exists an epidemic
threshold $\beta_{c}$, beyond which the density of infected nodes is
nonzero and increases as $\beta$ is increased. For
$\beta<\beta_{c}$, the epidemic goes extinct and $\rho=0$.
Figure~\ref{fig3} shows the dependence of the epidemic threshold
$\beta_{c}$ on the exponent $\gamma$ for different values of the
average degree $\langle k\rangle$ of the network. One can observe a
nonmonotonic behavior. For different values of $\langle k\rangle$,
$\beta_{c}$ is minimized for $\gamma \approx 2.2$.

\begin{figure}
\begin{center}
\scalebox{0.45}[0.45]{\includegraphics{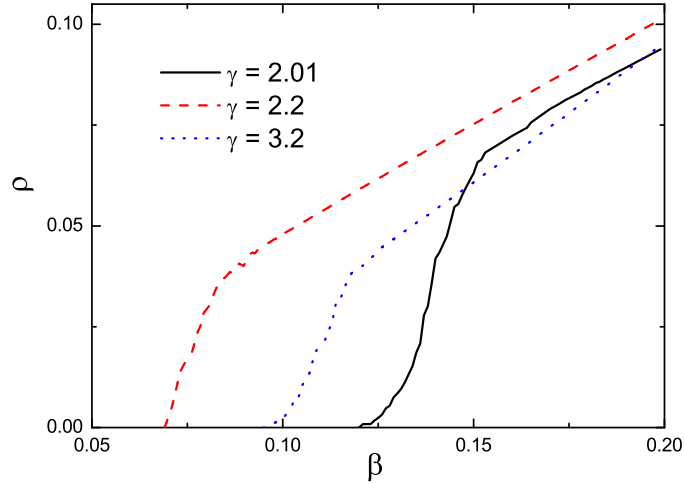}} \caption{ (Color
online) Density of infected nodes $\rho$ as a function of the
spreading rate $\beta$ for different values of the exponent
$\gamma$. The average degree of the network $\langle k \rangle=8$.
Each curve is an average of 30 different realizations.} \label{fig2}
\end{center}
\end{figure}

\begin{figure}
\begin{center}
\scalebox{0.45}[0.45]{\includegraphics{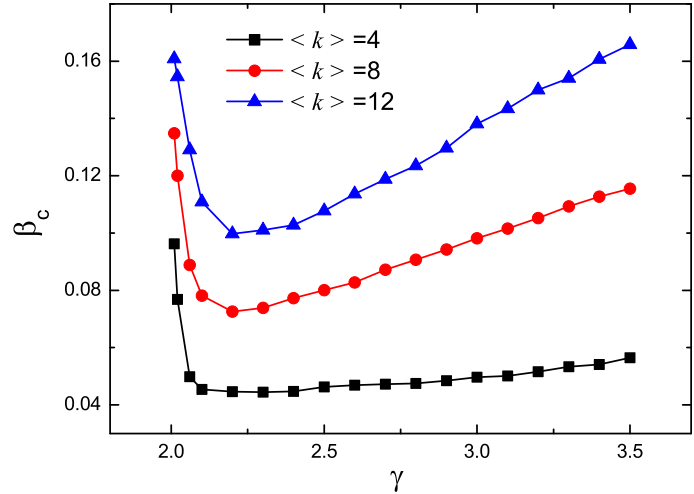}} \caption{(Color
online) The epidemic threshold $\beta_{c}$ as a function of the
exponent $\gamma$ for different values of $\langle k \rangle$. Each
data point results from an average over 30 different realizations.}
\label{fig3}
\end{center}
\end{figure}

\begin{figure}
\begin{center}
\scalebox{0.43}[0.43]{\includegraphics{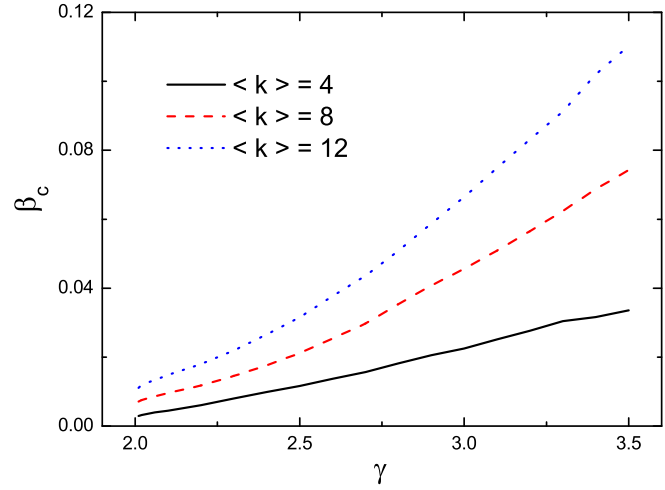}} \caption{(Color
online) The theoretical prediction of $\beta_{c}$ as a function of
$\gamma$ for different values of the average degree $\langle k
\rangle$ of a network. Each curve results from an average over 30
different realizations.} \label{fig4}
\end{center}
\end{figure}

According to the analysis of Ref.~\cite{Meloni}, the epidemic
threshold for uncorrelated networks is
\begin{equation}
\beta_{c}=\frac{\langle b_{\mathrm{alg}} \rangle}{\langle
b_{\mathrm{alg}}^{2} \rangle}\frac{1}{\lambda N},
\end{equation}
where $b_{\mathrm{alg}}$ is the algorithmic betweenness of a
node~\cite{alg1,alg2} and $\langle \cdot \rangle$ denotes the
average of all nodes. The algorithmic betweenness of a node is the
number of packets passing through that node when the
packet-generation rate $\lambda=1/N$~\cite{alg1,alg2}. For the
shortest-path routing protocol, the algorithmic betweenness is equal
to the topological betweenness ($b_{\mathrm{alg}}=b_{\mathrm{top}}$)
and $\langle b_{\mathrm{alg}} \rangle=\langle D \rangle/(N-1)$,
where $\langle D \rangle$ is the average topological distance of a
network. Here, the topological betweenness of a node $k$ is defined
as
\begin{equation}
b_{\mathrm{top}}^{k}=\frac{1}{N(N-1)}\sum_{i\neq
j}\frac{\sigma_{ij}(k)}{\sigma_{ij}},
\end{equation}
where $\sigma_{ij}$ is the total number of shortest paths going from
$i$ to $j$, and $\sigma_{ij}(k)$ is the number of shortest paths
going from $i$ to $j$ and passing through $k$. The average
topological distance of a network is given by $\langle D
\rangle=\sum_{i\neq j}d_{ij}/[N(N-1)]$, where $d_{ij}$ is the
shortest distance between $i$ and $j$. Combining Eq.~(3) and
Eq.~(4), we are able to calculate the theoretical value of the
epidemic threshold $\beta_{c}$. In Fig.~\ref{fig4}, one can notice
that for a given $\langle k \rangle$, the theoretical value of
$\beta_{c}$ increases with the exponent $\gamma$. However, in the
simulation results, $\beta_{c}$ decreases with $\gamma$ when
$\gamma<2.2$.

\begin{figure}
\begin{center}
\scalebox{0.53}[0.53]{\includegraphics{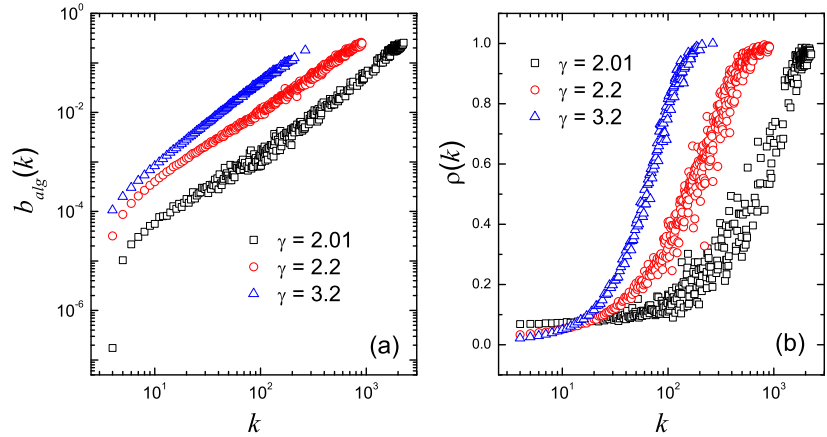}} \caption{(Color
online) (a) The algorithmic betweenness $b_{\mathrm{alg}}(k)$ and
(b) the density of infected nodes $\rho_{k}$ as a function of the
degree $k$ for different values of $\gamma$. For each value of
$\gamma$, we set the spreading rate $\beta=\beta_{c}+0.02$. The
average degree of the network $\langle k \rangle=8$. Each data point
results from an average over 30 different realizations.
}\label{fig5}
\end{center}
\end{figure}

To understand the deviation between numerical results and
theoretical analysis, we study traffic flow and infection
probability of the nodes with different degrees. From
Fig.~\ref{fig5}(a), one can see that the algorithmic betweenness
$b_{\mathrm{alg}}(k)$ increases with the degree $k$ and the
relationship between them follows a power-law form as
$b_{\mathrm{alg}}(k)\sim k\nu$ when $k$ is large.
Figure~\ref{fig5}(b) shows the density of infected nodes $\rho_{k}$
as a function of the degree $k$ for different values of $\gamma$.
One can see that $\rho_{k}$ increases with $k$. Combining Figs.
~\ref{fig5}(a) and (b), one can find that the algorithmic
betweenness is positively correlated with the probability of being
infected.

In the theoretical analysis, it is assumed that there is sufficient
number of nodes within each degree class $k$. However, when the
exponent $\gamma$ is very small (i.e., $\gamma<2.2$), the network
becomes highly heterogeneous and almost all nodes connect to the
initial $m$ nodes. As a result, these hubs carry almost all the
traffic flow and the epidemic threshold totally depends on only a
few hubs. Due to the uncertainty of infection, all the $m$ hubs may
simultaneously become susceptible when the spreading rate $\beta$ is
small. To make sure at least one hub is infected, the spreading rate
$\beta$ must be much higher than the theoretical prediction, leading
to deviation between numerical observations and theoretical
predictions of the epidemic threshold.

\section{Conclusion}\label{sec:conclusion}

In conclusion, we have studied traffic-driven epidemic spreading on
scale-free networks with tunable degree distribution. The
heterogeneity of networks decreases as the exponent $\gamma$ of the
power-law degree distribution increases. It is interesting to find
that the epidemic threshold is minimized at about $\gamma=2.2$.
Besides, we find that the nodes with larger degree have higher
traffic flow and thus are more likely to be infected. For
$\gamma>2.2$, both simulation results and theoretical analysis show
that the epidemic threshold increases with $\gamma$. For
$\gamma<2.2$, the network becomes so heterogeneous that the epidemic
threshold totally depends on only a few hubs. To ensure at least one
hub is infected, the spreading rate $\beta$ must be set to be a
relatively high value, leading to an enhancement of the epidemic
threshold. We hope our results can be useful to understand the
effect of network structures on traffic-driven epidemic spreading.

\section*{Acknowledgments}
This work was supported by the National Science Foundation of China
(Grant Nos. 61403083, 11275186, 91024026 and 71301028) and the
Natural Science Foundation of Fujian Province, China (Grant No.
2013J05007).


\begin{thebibliography}{0}

\bibitem{1} R. Pastor-Satorras, A. Vespignani,  Phys. Rev. Lett.
\textbf{86} (2001) 3200.

\bibitem{2}
M. E. J. Newman, Phys. Rev. E \textbf{66} (2002) 016128.

\bibitem{3}
M. Barth\'{e}lemy, A. Barrat, R. Pastor-Satorras, A. Vespignani,
Phys. Rev. Lett. \textbf{92} (2004) 178701


\bibitem{4}
G. Yan, Z.-Q. Fu, J. Ren, W.-X. Wang, Phys. Rev. E \textbf{75}
(2007) 016108.

\bibitem{5} M. Kitsak, L. K. Gallos, S. Havlin, F. Lijeros, L. Muchnik L, H. E.
Stanley, H. A. Makse, Nat. Phys. \textbf{6} (2010) 888.

\bibitem{6} R. Parshani, S. Carmi, S. Havlin, Phys. Rev. Lett. \textbf{104}
(2010) 258701.

\bibitem{7} C. Castellano, R. Pastor-Satorras, Phys. Rev. Lett. \textbf{105} (2010)
218701.

\bibitem{8} B. Karrer, M. E. J. Newman, Phys. Rev. E \textbf{84} (2011) 036106.
\bibitem{9} C. Castellano, R. Pastor-Satorras, Sci. Rep. \textbf{2} (2012)
372.
\bibitem{10} M. Dickison, S. Havlin, H. E. Stanley, Phys. Rev. E \textbf{85} (2012) 066109.
\bibitem{11} W. Wang, M. Tang, H. Yang, Y.-H. Do, Y.-C. Lai, G.-W.
Lee, Sci. Rep. \textbf{4} (2014) 5097.
\bibitem{12} J. Shang, L. Liu, X. Li, F. Xie, C. Wu, Physica A \textbf{419}
(2015) 171.


\bibitem{13} P. Echenique, J. G\'{o}mez-Garde\~{n}es, Y. Moreno, Phys. Rev. E \textbf{70} (2004)
056105.
\bibitem{14} W.-X. Wang, B.-H. Wang, C.-Y. Yin, Y.-B. Xie, T. Zhou, Phys.
Rev. E \textbf{73} (2006) 026111.
\bibitem{15} W.-X. Wang, C.-Y. Yin, G. Yan, B.-H. Wang, Phys. Rev. E \textbf{74}
(2006) 016101.
\bibitem{16} S. Meloni, J. G\'{o}mez-Garde\~{n}es, V. Latora, Y. Moreno, Phys. Rev. Lett. \textbf{100} (2008)
208701.
\bibitem{17} Z.-X. Wu, W.-X. Wang, K.-H. Yeung, New J. Phys \textbf{10} (2008) 023025.
\bibitem{18} M. Tang, Z. Liu, X. Liang, P. M. Hui, Phys. Rev. E \textbf{80} (2009) 026114.
\bibitem{19} H.-X. Yang, W.-X. Wang, Y.-B. Xie, Y.-C. Lai, B.-H. Wang, Phys. Rev. E \textbf{83} (2011) 016102.

\bibitem{19.1} W.-B. Du, Z.-X. Wu, K.-Q. Cai, Physica A \textbf{392}
(2013) 3505.
\bibitem{20} W. Huang, X. Yang, X. Yang, S. Chen, Physica A \textbf{410}
(2014) 22.
\bibitem{21} H.-X. Yang, M. Tang, Physica A \textbf{402}
(2014) 1.
\bibitem{22} C. Liu, W.-B. Du, W.-X. Wang, PLoS ONE \textbf{9} (2014)
e97822.


\bibitem{m1} V. Colizza, A. Barrat, M. Barth\'{e}lemy, A. Vespignani, Proc.
Natl Acad. Sci. USA \textbf{103} (2006) 2015.
\bibitem{m2} V. Colizza, A. Vespignani, Phys. Rev. Lett. \textbf{99} (2007)
148701.
\bibitem{m3} M. Tang, L. Liu, Z. Liu, Phys. Rev. E \textbf{79} (2009) 016108.

\bibitem{m4} D. Balcan, A. Vespignani, Nat. Phys. \textbf{7} (2011)
581.
\bibitem{m5} Z. Ruan, P. Hui, H. Lin, Z. Liu, Eur. Phys. J. B \textbf{86} (2013)
13.
\bibitem{m6} Y.-W. Gong, Y.-R. Song, G.-P. Jiang, Physica A \textbf{416} (2014)
208.

\bibitem{Meloni} S. Meloni, A. Arena, Y. Moreno, Proc. Natl Acad. Sci. USA
\textbf{106} (2009) 16897.




\bibitem{avoid} S. Meloni, N. Perra, A. Arenas, S. G\'{o}mez, Y.
Moreno, A. Vespignani, Sci. Rep. \textbf{1} (2011) 62.
\bibitem{yang1} H.-X. Yang, W.-X. Wang, Y.-C. Lai, Y.-B. Xie, B.-H. Wang, Phys.
Rev. E \textbf{84} (2011) 045101.
\bibitem{yang2} H.-X. Yang, W.-X. Wang, Y.-C. Lai, B.-H. Wang,  EPL \textbf{98} (2012)
68003.
\bibitem{yang3} H.-X. Yang, Z.-X. Wu,  J. Stat. Mech. (2014) P03018
\bibitem{yang4} H.-X. Yang, W.-X. Wang, Y.-C. Lai, Chaos \textbf{22} (2012) 043146.
\bibitem{yang5} H.-X. Yang, Z.-X. Wu, B.-H. Wang, Phys.
Rev. E \textbf{87} (2013) 064801.

\bibitem{exponent} R. Albert, A.-L. Barab\'{a}si,  Rev. Mod. Phys. \textbf{74} (2002)
47.
\bibitem{opinion1} V. Sood, S. Redner, Phys. Rev. Lett. \textbf{94} (2005) 178701.
\bibitem{opinion2} J.-S. Yang, I.-M. Kim, W. Kwak, 2009 EPL 88
20009.

\bibitem{yang6} H.-X. Yang, Z.-X. Wu, W.-B. Du, EPL \textbf{99} (2012) 10006.

\bibitem{Dorogovtsev} S. N. Dorogovtsev, J. F. F. Mendes, A. N. Samukhin, Phys. Rev. Lett. \textbf{85} (2000) 4633.
\bibitem{ba} A.-L. Barab\'{a}si, R. Albert, Science \textbf{286} (1999)
509.


\bibitem{SIS} N. T. J. Bailey, The Mathematical Theory of Infectious
Diseases, Griffin, London, 1975.

\bibitem{perc} M. Perc, New J. Phys. \textbf{11} (2009) 033027.


\bibitem{alg1} A. Arenas, A. D\'{\i}az-Guilera, R. Guimer\`{a}, Phys. Rev.
Lett. \textbf{86} (2001) 3196.

\bibitem{alg2} R. Guimer\`{a}, A. D\'{\i}az-Guilera, F. Vega-Redondo, A. Cabrales, A. Arenas, Phys. Rev. Lett. \textbf{89} (2002)
248701.


\end{thebibliography}
\end{document}